%% file: main.tex
\documentclass[sigconf,nonacm,authorversion]{acmart}

\AtBeginDocument{%
  }

\setcopyright{acmlicensed}
\copyrightyear{2025} 
\acmYear{2025} 
\setcopyright{cc}
\setcctype{by}
\acmConference[ETRA '25]{2025 Symposium on Eye Tracking Research and Applications}{May 26--29, 2025}{Tokyo, Japan}
\acmBooktitle{2025 Symposium on Eye Tracking Research and Applications (ETRA '25), May 26--29, 2025, Tokyo, Japan}\acmDOI{10.1145/3715669.3727349}
\acmISBN{979-8-4007-1487-0/2025/05}

\usepackage{lipsum}
\usepackage{outlines}
\usepackage{soul}
\usepackage{gensymb}
\usepackage{multicol,caption}
\usepackage{enumitem}
\usepackage{float}
\usepackage{dblfloatfix}    

\begin{document}

\title{Eye Gaze as a Signal for Conveying User Attention in Contextual AI Systems}

\author{Ethan Wilson}
\email{ethanrwilson1998@gmail.com}
\affiliation{%
  \institution{Meta Reality Labs Research}
  \city{Redmond}
  \state{Washington}
  \country{USA}
}

\author{Naveen Sendhilnathan}
\email{naveensn@meta.com}
\affiliation{%
  \institution{Meta Reality Labs Research}
  \city{Redmond}
  \state{Washington}
  \country{USA}
}

\author{Charlie S. Burlingham}
\email{cburlingham@meta.com}
\affiliation{%
  \institution{Meta Reality Labs Research}
  \city{Redmond}
  \state{Washington}
  \country{USA}
}

\author{Yusuf Mansour}
\email{ymans@meta.com}
\affiliation{%
  \institution{Meta Reality Labs Research}
  \city{Redmond}
  \state{Washington}
  \country{USA}
}

\author{Robert Cavin}
\email{robcavin@meta.com}
\affiliation{%
  \institution{Meta Reality Labs Research}
  \city{Redmond}
  \state{Washington}
  \country{USA}
}

\author{Sai Deep Tetali}
\email{saideept@meta.com}
\affiliation{%
  \institution{Meta Reality Labs}
  \city{Burlingame}
  \state{California}
  \country{USA}
}

\author{Ajoy Savio Fernandes}
\email{ajoyferns@meta.com}
\affiliation{%
  \institution{Meta Reality Labs Research}
  \city{Redmond}
  \state{Washington}
  \country{USA}
}

\author{Michael J. Proulx}
\email{michaelproulx@meta.com}
\affiliation{%
  \institution{Meta Reality Labs Research}
  \city{Redmond}
  \state{Washington}
  \country{USA}
}

\renewcommand{\shortauthors}{Wilson et al.}

\input{sections/abstract}




\maketitle


\input{sections/introduction}

\input{sections/related_literature}
\input{sections/eval_object_size}
\input{sections/eval_vlm}
\input{sections/discussion}


\bibliographystyle{ACM-Reference-Format}
\bibliography{biblio_baggins}

\end{document}

%% file: sections/abstract.tex
\begin{abstract}


Advanced multimodal AI agents can now collaborate with users to solve challenges in the world.  Yet, these emerging contextual AI systems rely on explicit communication channels between the user and system.  We hypothesize that implicit communication of the user's interests and intent would reduce friction and improve user experience when collaborating with AI agents.  In this work, we explore the potential of wearable eye tracking to convey signals about user attention.  We measure the eye tracking signal quality requirements to effectively map gaze traces to physical objects, then conduct experiments that provide visual scanpath history as additional context when querying vision language models.  Our results show that eye tracking provides high value as a user attention signal and can convey important context about the user's current task and interests, improving understanding of contextual AI agents.

\end{abstract}

%% file: sections/introduction.tex
\section{Introduction}

Artificial intelligence (AI) agents have become more connected with users in daily life~\cite{wienrich_extended_2021}, especially by observing context about the user's prior actions or current world state~\cite{zhang_vision_survey_2024}.  New innovations, such as vision-language models (VLMs)~\cite{li_foundation_2024} and machine perception devices~\cite{engel_aria_2023}, pave the way towards contextual AI agents: agents which ``\textit{see}" the nearby physical world and collect / compile contextual cues to better assist users.  Yet, current models interpret information differently than humans, so often misinterpret context, conflicting with user intent.  If implicit information about the user's state could be reliably supplied to the agent, the user and agent's intent could be better aligned.

Eye gaze has been hypothesized as a valuable signal for conveying intent to agents~\cite{ajanki_contextual_2010, buschel_here_2018, burlingham_scanpaths_2024, zhang_agi_2024}.  Gaze conveys information about objects users are interested in, cognitive load, the current action, etc.~\cite{mahanama2022eye}, all of which could improve models' understanding.  While eye tracking (ET) is a common input in extended reality (XR) systems~\cite{plopski_survey_2022}, the use of ET in human-agent interactions has only been lightly explored~\cite{sendhilnathan2024implicitgazeresearchxr}.  ET signals could convey to an agent what the user is or has been interested in.  Yet, wearable eye trackers are limited in accuracy due to multiple factors (system error, slippage, individual user differences, etc.)~\cite{ehinger2019new}, constraining whether objects could be reliably identified.  If an object's visual size relative to the ET signal accuracy is too small, it could be unreliable to detect.  

We present an analysis of the requirements and benefits of ET in wearable contextual AI.  Using a dataset of egocentric recordings taken during daily household tasks~\cite{pan_adt_2023}, we estimate the expected ET accuracy thresholds for detecting physical objects in different contexts.  We then conduct a number of experiments where contextual information from ET is appended to VLM queries.  These experiments reinforce ET's value in this space, and show improvements in the model's ability to perceive user attention and current actions.

\subsection{Contributions}

\begin{enumerate}[topsep=0pt]
    \item We estimate ET accuracy requirements needed for accurate gaze placement on physical objects, to determine ET accuracy requirements for wearable contextual AI and future systems.
    \item We explore how ET information can be conveyed to AI agents, both as point-in-time and as scanpath information with temporal dependencies.  By augmenting VLM queries by supplying ET context, we augment agentic ability to understand user attention and current actions.
\end{enumerate}

\subsection{Privacy and Ethics Statement}  Our findings convey ET's usefulness in human-agent interactions.  Eye movements are known to convey personal information and user preferences~\cite{bozkir2023eyetrackedvirtualrealitycomprehensive}, so any contextual AI system incorporating ET must be secure and privacy-preserving to avoid revealing user characteristics to others.

%% file: sections/related_literature.tex
\section{Related Literature}

Eye tracking is being adopted heavily in XR, providing clear value in human-computer interaction (HCI) interfaces.  As contextual AI emerges, new prototypes have explored eye gaze as a means to estimate user attention.  This section provides an overview of related literature, including the use of ET for selection, scene understanding, and ET in contextual AI.

\subsection{Eye Tracking for Selection in Extended Reality}

In recent years, eye gaze has gained popularity as a signal for HCI in XR systems~\cite{plopski_survey_2022}.  Eye gaze has comparable usability to controllers~\cite{luro_comparative_2019, zhang_telepresence_2019,fernandes2024effect} while freeing the hands for other tasks and being preferable to users~\cite{piening_looking_2021}.  While ET is prone to a \textit{Midas touch} fallacy, where false selections are made during ambient fixations~\cite{jacob_advanced_1995}, novel HCI methodologies~\cite{khamis_vrpursuits_2018} overcome this and make ET an ideal signal for interface navigation. Our work explores ET's ability to continuously and implicitly convey a user's attention and intent~\cite{sendhilnathan2024implicitgazeresearchxr}.  This has only lightly been explored with AI agents, though ET has facilitated automatic contextual displays in the past~\cite{toyama_recognition_2012}.

\subsection{Eye Gaze Encodes Scene Understanding}

Eye movements reflect a viewer's internal processing of a scene, giving insights to cognitive state and attention as one interprets new visual stimuli~\cite{langton_eyes_2000, eckstein_beyond_2017}. Sequences of gaze fixations (i.e., scanpaths) encodes contextual cues as to future objects of interest~\cite{itti_computational_2001, burlingham_laziness_2024}; a number of works have leveraged scanpath history for short-term gaze prediction / anticipation~\cite{huang_predicting_2018, david-john_prediction_2021, Hu2021, damelio_tppgaze_2024}. Burlingham et al. found temporal dependencies in scanpaths lasting for 4-5 fixations on average, with high variance across task types~\cite{burlingham_scanpaths_2024}. Contextual AI models may be able to leverage the rich, multiscale structure of scanpaths when inferring intent. 

Insights about the cognitive encoding of nearby objects can inform our expectations for eye movements in contextual AI~\cite{Tatler2011}.  For example, humans tend to look at a coffee mug just before grasping, to encode the location of the handle so that it can be successfully grasped.  Object locations are encoded via an egocentric reference to the user, and greater affordances are given to nearby interactable objects~\cite{costantini_affordance_2010, Tatler2011}.  The visual system elicits responses to reachable 3D objects, even when there is no intent to interact~\cite{iachini_motor_2014, iachini_temporal_2023}.  So, by analyzing the eye gaze fixations on nearby objects, we could predict possible future interactions. 

\subsection{Eye Tracking in Contextual AI}

Information from the physical world could greatly improve user interactions with AI agents~\cite{zhang_vision_survey_2024}.  Emerging products, such as the Ray-Ban Meta\footnote{\url{https://www.meta.com/smart-glasses}} and Google's Ask Photos\footnote{\url{https://blog.google/products/photos/ask-photos-google-io-2024/}}, use image context to improve user interaction.  But, given a full image, the human's intent may not align with the salient features detected by the agent.  Eye gaze could aid in narrowing relevant context observed by contextual agents~\cite{ajanki_contextual_2010, buschel_here_2018}, enhancing understanding and avoiding hallucination~\cite{cui_holistic_2023, leng_hallucinations_2024}.  

Some wearable contextual AI prototypes integrating eye tracking have been proposed~\cite{zhang_agi_2024}.  The GazeGPT system projects 2D gaze onto an image capture, cropping image contents before interfacing with a VLM~\cite{konrad_gazegpt_2024}.  They demonstrated gaze-based querying to be faster, more accurate, and more natural than head-mounted and smart-phone-like baselines.  G-VOILA interfaces with a textual LLM, using gaze-generated saliency maps for prompt adjustment~\cite{wang_gvoila_2024}.  Derived object information is spliced into the query, increasing robustness against ambiguity and increasing participants' confidence in the system.  These prototypes show clear value from the inclusion of ET for point-in-time querying.  

%% file: sections/eval_object_size.tex
\section{Evaluation of Eye Tracking Signal Quality Requirements}

To better understand ET's role in future contextual AI systems, we first estimate the ET signal quality needed for accurate gaze-based selection of physical objects.  Contextual AI models that collaborate with users would benefit from knowledge of users' real-world interests.  ET is a promising signal to capture the user's attention, both at one point in time~\cite{konrad_gazegpt_2024}, or continuously to build historical context.  We hypothesize that the visual angle subtended by objects that users ``look at'' defines a lower bound on ET signal accuracy.  By measuring the visual angle expressed by nearby objects, we can approximate the ET accuracy required to consistently track a user's point of focus.

To investigate accuracy requirements, we analyze objects that are nearby in the user's field of view (FOV) during daily household tasks, then relate the object size statistics to ET signal quality requirements.  Because we expect humans and AI agents to collaborate in daily life, it is important for the ET system to achieve an accuracy which allows consistent, accurate attention modeling in a broad set of scenarios~\cite{feit2017toward}.  Individuals are far more likely to look at or interact with objects in the immediate vicinity~\cite{ballendat_proxemic_2010}, so we constrain this analysis to objects which are nearby candidates for interaction.

\subsection{Dataset}

For this analyses, we use a subset of the Aria Digital Twins (ADT) dataset~\cite{pan_adt_2023}.  The ADT dataset contains egocentric recordings of daily-life tasks in indoor environments.  We analyze the recordings in the furnished apartment scene where one user performs tasks (we omit multiple-user recordings to better focus on human-object interaction rather than social interactions), totaling 93 recordings and $\sim$3 hours of footage.  Designed to model real-world household scenarios, these recordings span the following tasks:  decorating, cooking, working, cleaning, and object examination.

In addition to the ET signal (median error $= 1.5\degree$) provided by Project Aria glasses~\cite{engel_aria_2023}, the ADT dataset contains ground-truth information about physical objects in the scene.  Each object's position, orientation, bounding box, and segmentation region is tracked throughout the recording, with median tracking error of $5 mm$.  This ground-truth data enables the analysis of object visual statistics at a fine scale, detection of human-object interactions, and accurate placement of gaze on objects.  There are 396 distinct household objects tracked in the dataset, with varying presence across the different tasks and recordings.

\subsection{Object Visual Size in Relation to Eye Tracking Error}

ET spatial error is the measured bias between the ground truth and estimated gaze positions.  This error persists following techniques such as ET calibration and fixation detection~\cite{schuetz_eye_2022}.  We measure spatial error as an angular offset between the user's true gaze point and computed value.  We wish to approximate the influence of ET spatial error in placing fixations on objects, to better inform how reliably an ET system could convey a user's attention on world objects.

\begin{figure}[t]
 \centering
 \includegraphics[width=\linewidth]{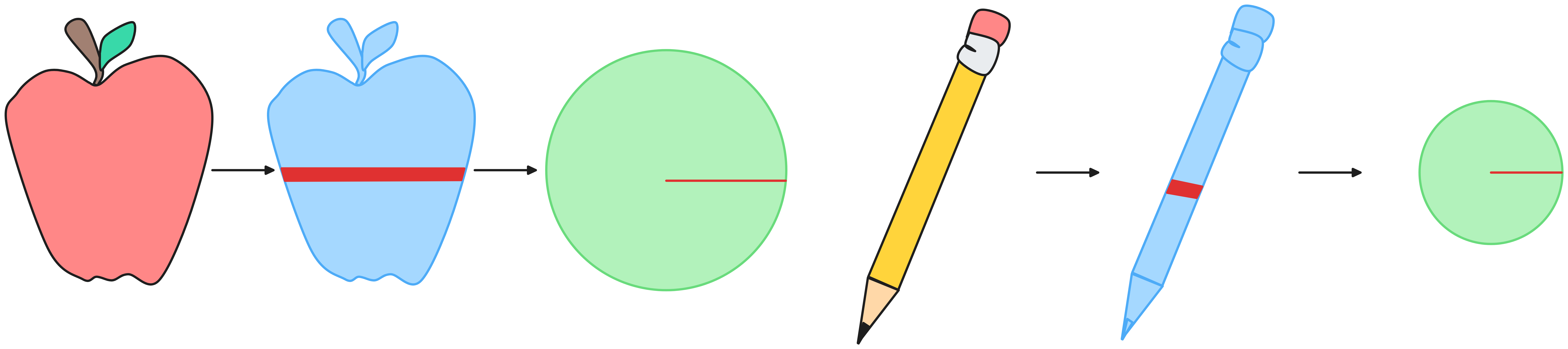}
 \caption{Illustration of eye tracking spatial error and object visual size measurements.  As an average case measurement, object segmentation area can be mapped to a circular region, with the radius reflecting the eye tracking accuracy requirement (\textcolor{red}{thin bars}).  Alternatively, $1 / 2$ minor axis span $L_{min}$ (\textcolor{red}{thick bars}) is a stricter bound for measuring non-uniform objects.}
 \label{fig:et_relationship}
\end{figure}

Our metrics to predict ET accuracy needs are derived from object visual size --- the visual angle spanned by the object relative to the user's FOV.  Visual size is related to both the physical size of an object and its distance from the user.  We model object visual size from the total segmentation area $A_{{}_{seg}}$ of an object in a linear camera model, which measures degrees$^2$ occupied by the object.  To convey visual size against a one-dimensional ET error requirement $err_{{}_{ET}}$, we convert visual size to the approximate \textbf{radius} of the object. $err_{{}_{ET}} \ \leq \ \sqrt{A_{{}_{seg}} / \pi}$.  This inequality approximates an \textbf{average case} for the ET error requirement, and is valid when objects have roughly uniform dimensions.  To account for non-uniform objects, we compute a more \textbf{conservative} ET error as $1 / 2$ the \textbf{minor axis} span $L_{min}$ of the object's segmentation region: $err_{{}_{ET}\ low} \ \leq \ 1 / 2 \ L_{min}$.  \autoref{fig:et_relationship} illustrates these metrics and the relationship between ET error requirements and object visual size.

\subsection{Protocol}

We approximate the ET requirements for daily use by observing the household tasks being performed in the ADT dataset~\cite{pan_adt_2023}.  We measure the distributions of object visual sizes for $err_{{}_{ET}}$ and $err_{{}_{ET}\ low}$.  To better inform various contextual AI applications, we specify the ET requirements across different interaction spaces, namely:
\begin{enumerate}  [topsep=0pt]
    \item \textbf{Near-field objects:} all objects within 1 meter of the participants.
    \item \textbf{Mid-field objects:} all objects between 1 - 2 meters of the participants.
    \item \textbf{Interacted objects:} all objects being physically interacted with (held, pressed, pushed, etc.) by the participants, with a start / stop padding of 1 second for the interaction.
    \item \textbf{Fixated objects:} all objects within 2 meters fixated on by participants' gaze as they navigate the scenes.
\end{enumerate}

\begin{figure*}[t]
    \centering
    \includegraphics[width=\linewidth]{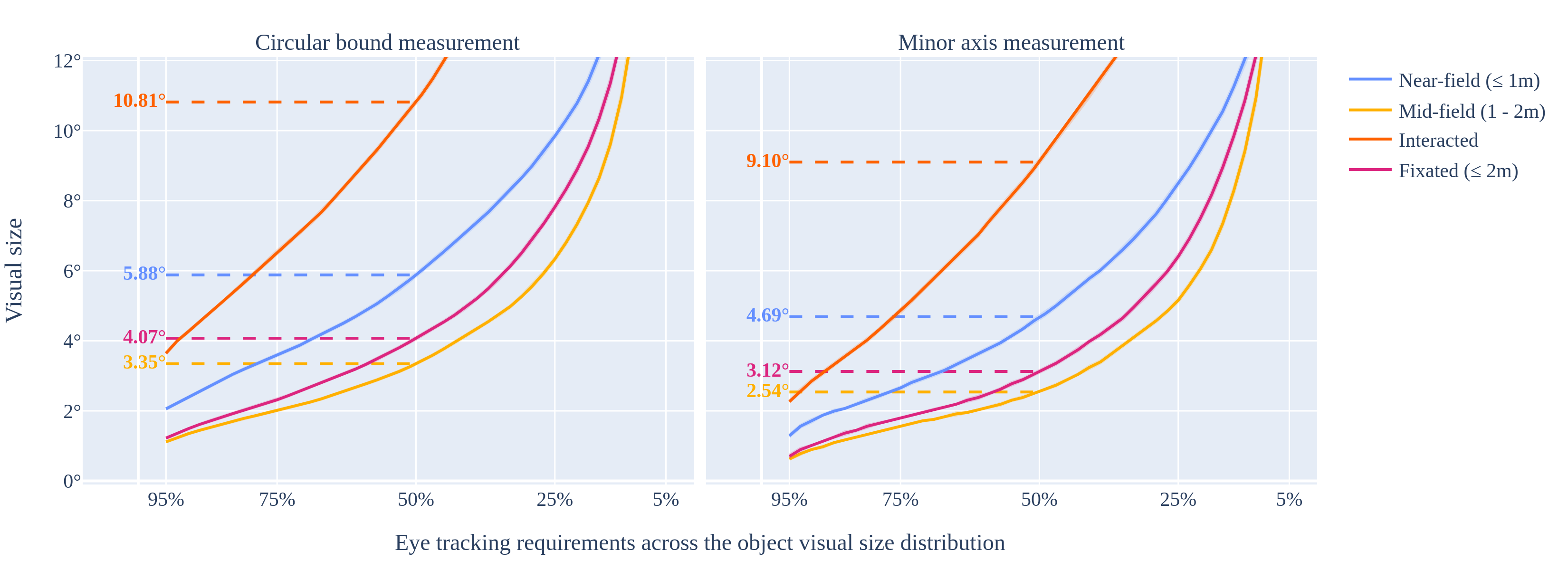}
    \caption{Eye tracking accuracy requirements to place gaze accurately within in the ADT dataset, where users performed household actions in an indoor environment~\cite{pan_adt_2023}  Near-field and mid-field measurements consider all objects present in the user's field of view.  Interacted objects are being actively manipulated by the user, and fixated objects consider those being directly gazed at.}
    \label{fig:visual_sizes}
\end{figure*}

\subsection{Results}
\label{sec:accuracy_results}

The \textbf{interacted} measurement aggregates object statistics when being manually interacted with, including picking up, pushing, pressing, etc. The \textbf{near-field} ($\leq$ 1 meter) and \textbf{mid-field} (1 - 2 meters) measurements reflect the visual FOV occupied by \textit{every} object in the environment within the distance threshold.  \textbf{Fixation} measurements consider objects that are within 2 meters of the participant at the time of fixation.  Considering ADT's household scenarios, the \textbf{interacted} and \textbf{fixation} categories reflect the distribution of objects likely to be of interest during daily tasks.

To estimate ET accuracy requirements, we compute the entire distribution of object visual sizes recorded in camera projection space. To place gaze on an object of average (projected) size 50\% of the time, we measure at the distribution's 50\% mark.  These measurements can help to inform system design; while 50\% reliability may be \textit{useful supporting context} in a broader contextual AI agent, a system relying heavily on ET may aim for higher coverage.  The distributions for each interaction scenario are seen in \autoref{fig:visual_sizes}.

Wearable ET accuracy is known to suffer in dynamic conditions~\cite{onkhar_tobii_2023}, yet recent devices remain quite accurate in unconstrained settings\footnote{\url{https://www.tobii.com/products/eye-trackers/wearables/tobii-pro-glasses-3}}\footnote{\url{https://pupil-labs.com/products/neon}}.  Assuming a device with $\leq 3 \degree$ accuracy during daily wear, our results indicate that the majority of fixated objects (radius average=$4.07\degree$; minor axis=$3.12\degree$), the majority of objects in the near-field (radius=$5.88\degree$; minor axis=$4.69\degree$), and nearly all interacted objects (radius=$10.81\degree$; minor axis=$9.10\degree$) are reliable for placing gaze on the correct object.  Conversely, objects in the mid-field (radius=$3.3\degree$; minor axis=$2.54\degree$) will be somewhat unreliable at this signal quality, where roughly half of objects are not able to be detected.

%% file: sections/eval_vlm.tex
\begin{figure*}[t]
    \centering
    \includegraphics[width=\linewidth]{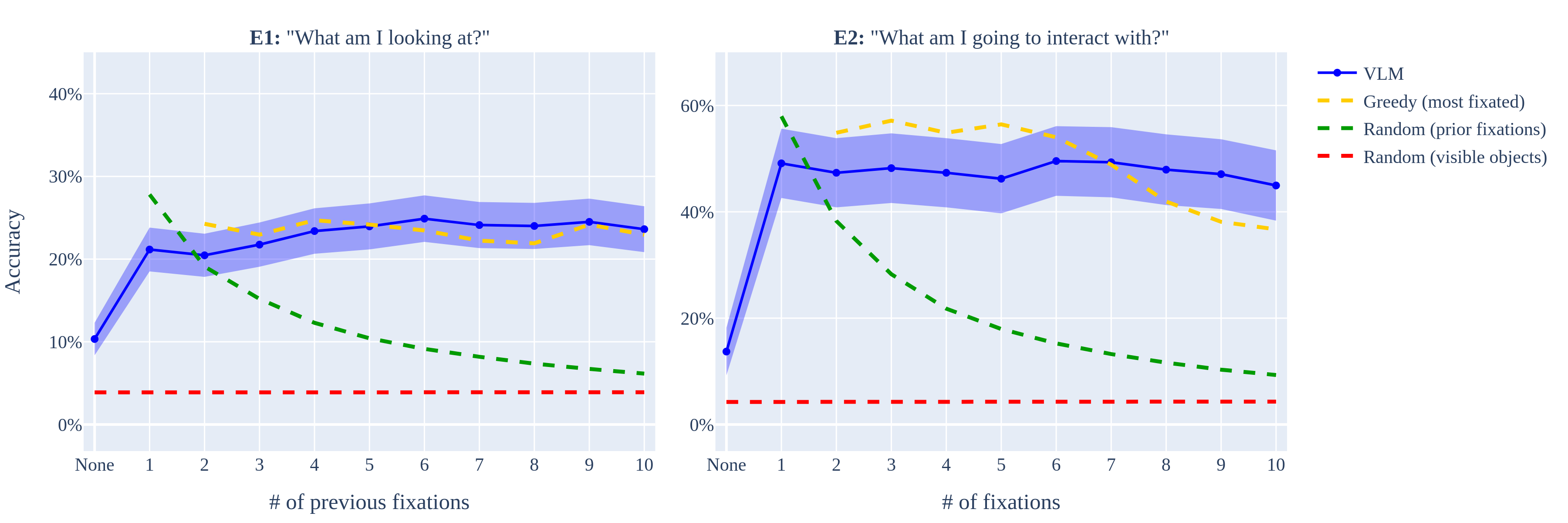}
    \caption{Experiments where prior gaze fixation contents are supplied to a VLM along with egocentric images.  When many fixations are given as context, the model can synthesize image + gaze information to outperform a greedy baseline that only considers contents from the prompt. The error surfaces in light blue represent 95\% confidence intervals.}
    \label{fig:e1e2}
\end{figure*}

\section{Eye Tracking Context in Vision-language Model Queries}

To begin to explore the value that ET signals can provide in contextual AI systems, we model experiments which reflect potential end-to-end contextual AI systems.  In these experiments, we build up historical context by creating timelines of past fixations on physical objects.  This context is included in VLM queries to measure positive impacts on model understanding.  Our baseline comparison is a VLM query which uses only the egocentric image as added context (similar to a Ray-Ban Meta or Google Ask Photos query).

\subsection{Methodology}

The Meta Llama 3.2 90B VLM\footnote{\url{https://ai.meta.com/blog/llama-3-2-connect-2024-vision-edge-mobile-devices/}}~\cite{grattafiori_llama_2024} is used as a contextual AI agent.  Queries consist of an egocentric image, a main query, and additional prompting to inject context.  In both experiments, the agent is constrained via JSON to respond with one option from all currently visible objects.  We are operating under the pretense that in a full system, an object recognition / scene understanding model would be available.  Note that a model tuned for egocentric image understanding and / or for a specific task would likely see improved results.  Yet, these experiments indicate the added value when incorporating ET contextual information.

\subsubsection{E1: ``What am I looking at?" with Historical Context}

\begin{figure*}[t]
    \centering
    \includegraphics[width=0.9\linewidth]{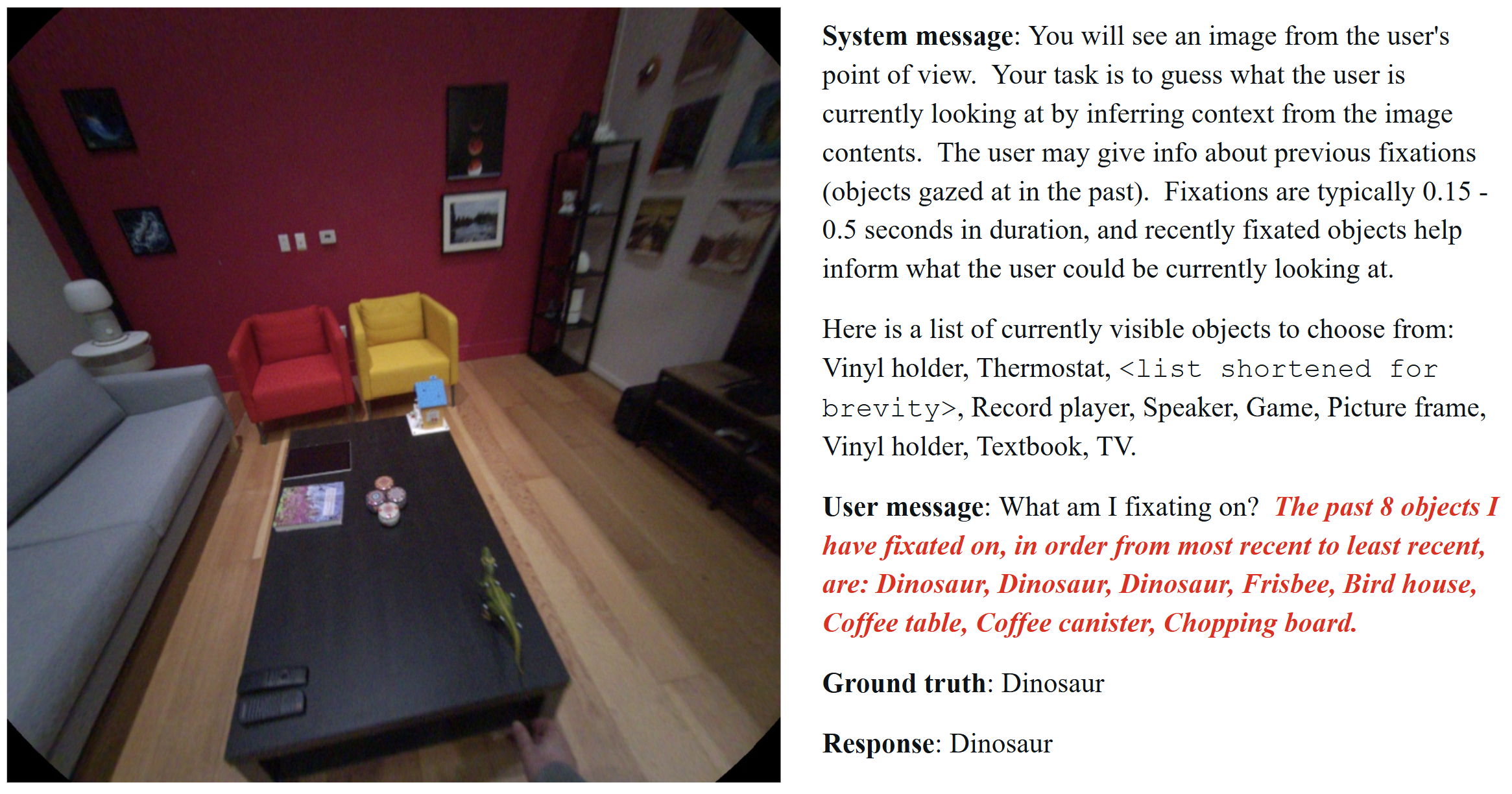}
    \caption{An example query to the VLM for E1.  The additional context from fixation history is highlighted in red; We vary the amount of context given to the model to measure how this context influences model understanding.}
    \label{fig:query_example}
\end{figure*}

In this experiment, we pose the question ``what am I looking at?"  This experiment serves as a benchmark for the effect that prior eye gaze context serves in improving image understanding.  We first detect and localize fixations on objects\footnote{We use a velocity-thresholding algorithm at 100$\degree$ per second~\cite{salvucci_fixations_2000}, to account for the relatively low sampling rate of Project Aria glasses (30Hz)~\cite{engel_aria_2023}.  We only consider fixations $\geq$ 150 ms for analysis.}, then perform uniform random sampling across each recording in the ADT dataset to analyze 919 image frames which each contain at least 10 prior fixations.  At each sample, we make multiple queries, varying the number of prior fixations supplied as context to the VLM, between 0 - 10.  An example prompt set can be seen in~\autoref{fig:query_example}.

\subsubsection{E2: "What am I going to interact with?" with Historical Context}

We query the VLM ``what am I going to interact with?" while again varying the amount of gaze context.  In this experiment, we supply the image from a current fixation, where a physical interaction is guaranteed to occur within the next second and at least 10 prior fixations exist (237 distinct image frames).  The types of physical interactions detected in ADT are grasping, pushing, and pulling with hands.  This task models the use of ET as supporting context for user action understanding and prediction.

\subsection{Results}

Because the VLM model is constrained to respond by selecting from the list of all currently visible objects, these experiments are classification tasks where \textit{accuracy = correct selections / all trials}.  Cases where the VLM response failed to return parseable JSON ($<$1\% of trials) were discarded.

A number of baselines are compared against to see if the model effectively uses the context supplied.  These baselines implement simple heuristics on the image and gaze context.  The lowest performing baseline is random guessing among all visible objects.  We also implement random guesses from the list of previously fixated objects, and a greedy strategy to always choose the most fixated object.  Note these baselines still use context provided from ET, but make no attempt to synthesize with the egocentric image for better world understanding.

\subsubsection{E1}

When querying the VLM only with the egocentric image as context, the model successfully predicts the current fixated object 10.3\% (95\% CI = [8.3\%, 12.3\%]) of the time (see \autoref{fig:e1e2} (left)).  An effective strategy is to always return the immediately preceding fixation (left tail of Random (prior fixations) in \autoref{fig:e1e2}).  VLMs are not expected to excel at gaze prediction, as they are known to misinterpret context or hallucinate~\cite{cui_holistic_2023, leng_hallucinations_2024}.  However, including context from prior gaze greatly improves the model's ability to predict the current fixation, with a peak accuracy of 24.8\% (CI = [22.1\%, 27.7\%]) at 6 prior fixations.  The gaze context-based baselines slightly outperform the VLM with one or few prior fixations, reinforcing that current gaze is contingent on scanpath history~\cite{burlingham_laziness_2024, burlingham_scanpaths_2024}.  With more context (6+ fixations), the model begins to outperforms baselines, indicating that prior context and image contents are being synthesized, and the combination of contextual cues increases the model's performance.

\subsubsection{E2}

E2 sees similar trends to E1; however, the more contextually-grounded task of action prediction sees greater benefit from ET context. Clearly, prior eye gaze is a strong indicator for interaction, and historical gaze could greatly improve the VLM's ability to understand the user's actions. Peak accuracy is 49.5\% (CI = [43\%, 56.1\%]). As evident by this and the stronger baseline performances, gaze is tightly coupled with the onset of interaction. Note that queries all are positive examples where an interaction does take place, and the inclusion of a null case could have led to the model raising false positives / negatives.

%% file: sections/discussion.tex
\section{Discussion}



We expect that ET's value would become even more prominent in future models which are trained specifically for egocentric understanding and / or with eye gaze as a direct input~\cite{koorathota_attention_2023}.  Our findings, building on prior works~\cite{toyama_recognition_2012,  burlingham_scanpaths_2024}, evidence that human actions and gaze patterns display temporal dependencies contingent with previous actions, similar to the dependencies in written language.  If we can effectively convey the traces of human attention and actions, VLMs may become able to better infer current / future context based on the patterns present in prior behavior.

Our ET signal quality benchmark measures the likelihood of sensing objects, but has little considerations of edge cases (such as very small or very far objects).  In the future, supplementary computations might aid the ability to place gaze on the correct object, possibly via contextual cues~\cite{bi_bayesian_2013} for error correction or additional sensors~\cite{wei_predicting_2023}.  These could alleviate ET sensor quality from becoming a bottleneck when using gaze to infer human attention, specifically in more challenging contexts such as outdoors, where objects tend to be much further.

\subsection{Future Work}

While the ADT dataset provided a platform for simulating gaze-aided querying~\cite{pan_adt_2023}, it is critical to explore the impacts of eye tracking in real contextual AI scenarios.  Prototypes leveraging point-in-time gaze have seen high user acceptance~\cite{konrad_gazegpt_2024, wang_gvoila_2024}, and the inclusion of temporal context is likely to better improve an agent's ability to infer context and disambiguate a user's queries.  Thus, user experience investigations are important future avenues to follow up.

This work explored contextual inferences that could be made in short intervals lasting no longer than a few seconds, yet the information contained in larger time scales could also be invaluable for improving contextual AI agent understanding.  Larger scales could enable better inferences of the current situation, and pave the way towards implicit personalization of contextual AI assistants~\cite{pardini_role_2022}.

\subsection{Conclusion}

We investigated eye tracking signals' ability to improve multimodal agents' understanding of the physical world.  Our results suggest that for close by scenarios, such as active grabbing / touching of objects and gaze selection, current ET systems could consistently place fixations on objects and convey relevant information to VLM agents.  In our experiments, we saw direct benefits when adding scanpath history to queries. Given these findings, future contextual agents which receive signals about user attentive state may obtain a greater understanding about the world, and better align with user intent, improving the usability of such systems.
